\documentclass[12pt,a4paper]{article}
\usepackage{amsmath}
\usepackage{amsfonts}
\usepackage{amssymb}
\usepackage{color}

\usepackage{latexsym}
\usepackage{pstricks}
\usepackage{latexsym}  
\usepackage{epsfig}
\usepackage{graphics}
\usepackage{color}
\def\half{\frac 1 2}
\def\x{\mathbf{x}}
\def\y{\mathbf{y}}
\def\ba {\begin{eqnarray}}
\def\ea {\end{eqnarray}}
\def\be{\begin{equation}}
\def\ee{\end{equation}}
\newcommand{\nn}{\nonumber}
\title{Conformal Field Theory in $D>2$ dimensions,\\
representations and harmonic analysis}
\author{Gerhard Mack \\ $\quad$\\
II. Institut f\"ur Theoretische Physik\\
der Universit\"at Hamburg\\[1mm] 
talk presented at the conference on \\ Representation Theory and Harmonic Analysis \\ at
Paris Saclay 5 December 2018}

\date{} 

\begin{document}

\maketitle
%
\section{Plan of the talk}
\begin{itemize}
\item
Summary of what happend since 1968\\
(when I first met Ivan Todorov)
\item
The relation between representation theory and harmonic analysis 
for groups $SO(D,2)$ and $SO(D+1,1)$\\ $\   $ i.e. a group theoretical
 version of the familiar relation between field theory in Minkowski and Euklidean space
\item recent developments: Are we close to the {\em analytic} solution of the 3-dimensional 
(conformal) Ising model?
\end{itemize}
%
The second point is important: positions and residues of poles in the 
 partial wave amplitude $g(\chi=[l,\delta])$ which is obtained by partial wave analysis of the {\bf Euklidean} 4-point Green functions are restricted by the requirement that they are associated with unitary positive energy ray representations of
 the {\bf Minkowskian} conformal group $SO(D,2)$. 

And it can also be a source of interesting mathematics.
\section{What happened since 1968? 
}
I met Ivan Todorov in 1968 at the ICTP Trieste.
I had been invited by Abdus Salam, who had studied my PhD-thesis on dilatation- and conformal symmetry \cite{Mack0} of February 1967.The collaboration with Abdus Salam resulted in a joined paper \cite{MackSalam}.  Abdus Salam also pointed out Dirac's early work on conformal symmetry \cite{Dirac} which introduced the conformal covariant description of ($D=4$-dimensional) space time using lightlike $D+2$ - vectors $\xi$

Ivan had worked at the  research center in Dubna in the USSR, which also hosted scientist from Bulgaria and other states of the Warsaw pact. It was Ivans
 first visit to the west.
We began our collaboration on conformal symmetry with a paper \cite{MackTodorov}
 on the ladder reprentation of the conformal group $SO(4,2)$. This is a unitary positive energy representation 
 which also serves to label the states of the $H$-atom.

We agreed to meet again at the Vienna conference on High Energy physics that year. Shortly before the conference, the Soviet Union's troops invaded 
Czechoslovakia, and I got worried whether Ivan would come. But never before 
had there been so many scientists from eastern Europe at a conference in the
 west.
 I met Ivan and our paper got finished and published.

This was the beginning of a collaboration lasting 25 years. Our next paper demonstrated freedom of Conformal invariant Green functions from
 ultraviolet divergences \cite{MackTodorovUV}

Wilsons 1969 paper on operator product expansions (OPE) \cite{Wilson} cited me honorably and created a burst of activities:\\
 Polyakov \cite{Polyakov} (1970),  Migdal \cite{Migdal}(1971), 
d'Eramo, Parisi and Peliti \cite{ParisiPeliti, DEramoParisiPeliti} (1971), 
Symanzik \cite{Symanzik, Symanzik1, MackSymanzik} (1971), Ferrara, Gatto and Grillo, \cite{FerraraGattoGrillo,FerraraGrilloGatto} (1972), same and Parisi \cite{FerraraGattoGrilloParisi} (1972), Tonin et al.\cite{BonoraCiccarielloSartoriTonin}, Brezin, Itzykson, Zuber  and Zinn Justin.

I began working on a group theoretical approach to conformal field theory in 
1972. It used group representation theory and harmonic analysis on groups in the spirit of Eugen Wigner \cite{Mack1,Mack2,Mack3}.

The work was completed by Dobrev, Petkova, Petrova and Todorov \cite{DobrevPetkovaPetrovaTodorov}
and published jointly as a book \cite{DobrevMackPetkovaPetrovaTodorov}:
{\em Harmonic Analysis
on the $n$-dimensional Lorentz Group and its Application to Conformal Quantum Field Theory}, Lecture Notes im Physics {\bf 63}, Springer 1977

I had met Eugen Wigner at the Universiy of Princeton. My PhD adviser Hans 
Kastrup had been invited by Wigner who had studied Kastrups papers on
 conformal symmetry \cite{KastrupAP,KastrupNPB} This visit resulted in further papers on the subject \cite{KastrupPR}. Hans Kastrup  paved the way to my visit. 
The visit was financed by the
 Studienstiftung des Deutschen Volkes.

To prepare, I had studied Wigners book on group theory in quantum mechanics. 
He became my hero and influences  my work until now

Eugen Wigner and his wife invited Arthur Jaffe and me to a long weekend in 
their winter home in Vermont. And I met him repeatedly again later on. At the conference of Nobel Laureates in Trieste which was organized by Abdus Salam. 
And during later visits in Princeton. The last encounter was at the conference
 in Baltimore which celebrated Wigners 90th birthday. Arthur Jaffe was also 
there. A greeting message by Abdus Salam was presented  in which he explained how much he had learned from Wigner. 

In this talk I will only speak about conformal field theory in more than 2 dimensions. A major breakthrough in 2-dimensional conformal field theory occurred in 1984 with the paper \cite{BelavinPolyakovZamolodchikov} of Belavin, Polyakov and Zamolodchikov on 
{\em Infinite conformal symmetry in two-dimensional quantum field theory}. L\"uscher 
and myself had discovered before \cite{mackUnpub} that the 2-dimensional traceless stress 
energy tensor is a Lie field; they also found the bound $c\geq 1/2$ on the coefficient of the central term. This paper remained unpublished, but I 
presented it at a seminar in Copenhagen which Polyakov attended.

\begin{figure}[t!]
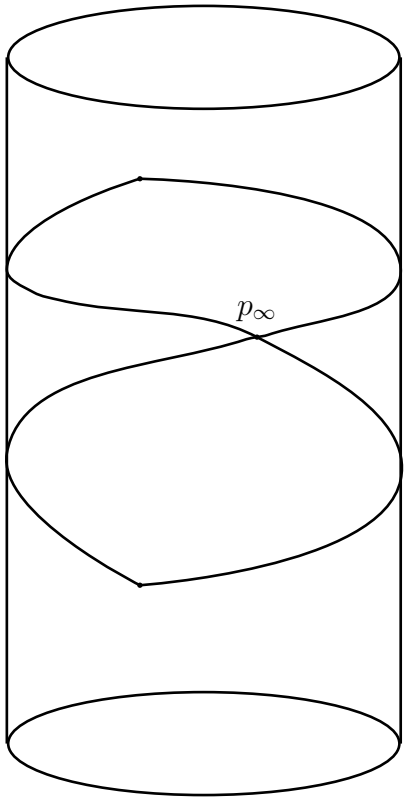

\psset{unit=0.7cm}

\psset{dotsize=0.1 0}
\psset{linewidth=.05}
\pspicture(0,0)(10,15.2)
\psline(0,1)(0,14)
\psline(7.4,1)(7.4,14)
\rput(4.7,9.2){$p_\infty$}
\psdot(2.5,4) 
\psdot(2.5,11.7)
\psdot(4.7,8.7) 
\pscurve(2.5,4)(0,6.5)(4.5,8.66)(4.7,8.7)(4.9,8.74)(7.4,10)(2.5,11.7) 
\pscurve(2.5,4)(7.4,6)(4.7,8.7)(1,9.4)(0.5,9.55)(0,10)(2.5,11.7)
%
\psellipse(3.7,1)(3.7,1)
\psellipse(3.7,14)(3.7,1)
\endpspicture
\caption{The universal covering $\tilde{M}$ of conformally compactified Minkowski space time is a tube with  point(s) $p_{\infty}$ at spatial infinity of Minkowski space located on the back of the tube. The boundary of Minkowski space passes through it. The two corner points of the boundary of Minkowski space are on the front of the tube.
 \label{fig:spaceTime}
}
\end{figure}

\section{The space time manifold $\tilde{M}$}
Conformal field theory satisfies locality \cite{LuescherMack}.

It lives on the infinite sheeted covering $\tilde{M}$ - a tube (Figure 1) -  of Minkowski space which admits a conformal invariant causal structure. Minkowski space is imbedded into $\tilde{M}$ and is characterized by its point(s) at spatial infinity
$p_\infty$.

\subsection{Lagrangean approach}
There exists a coupled set of integral equations for all the n-point functions 
of $\phi^3$-theory \cite{Symanzik}. They can also be used for $\phi^4$-theory by considering it as a theory with the fundamental fields $\phi(x)$ and $:\phi^2:(x)$.
The equations for propagator and 3-vertex require renormalization; 
renormalized versions have been derived in the 60's \cite{Symanzik}. The 4-point function furnishes also the Bethe Salpeter kernel.
\subsection{Bethe-Salpeter equations $\mapsto$ define kernels}
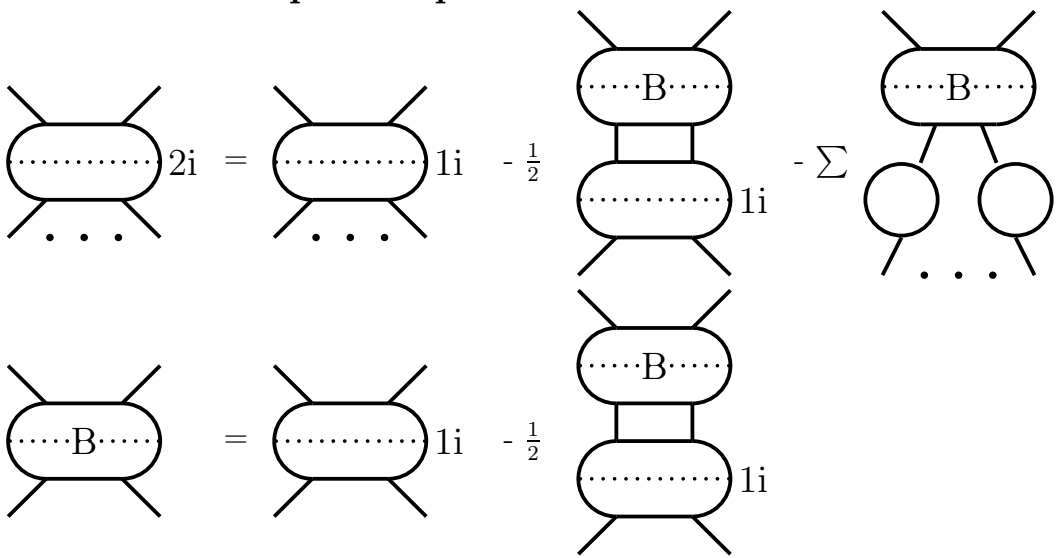
\psset{unit=1cm}
\psset{dotsize=0.2 0}
\psset{linewidth=.05}
\def\OvalShort { 
\psline(-.5,-.5)(.5, -.5)
\psline(-.5,.5)(.5,.5)
\psarc(-.5,0){.5}{90}{270}
\psarc(.5,0){.5}{270}{90}
}
\def\OvalLong { 
\psline(-.75,-.5)(.75, -.5)
\psline(-.75,.5)(.75,.5)
\psarc(-.75,0){.5}{90}{270}
\psarc(.75,0){.5}{270}{90}
}
\def\BSKern {
\psline(-.5,-.5)(.5, -.5)
\psline(-.5,.5)(.5,.5)
\psarc(-.5,0){.5}{90}{270}
\psarc(.5,0){.5}{270}{90}
\psline[linestyle=dotted](-1.,0)(1.,0)
}

\def\CircleUp{
\pscircle(0,0){.5}
\psline(0,.5)(0,1.5)
}
\def\CircleFeet{
\psline(-1.,-1.)(-0.3,-0.3)
\psline(1.,-1.)(0.3,-0.3)
\psline[linewidth=.1,linestyle=dotted,dotsep=.3](-.5,-1)(.5,-1)
}
\def\OvalShortTwoFeet {
\psline(-1.,-1.)(-0.5,-0.5)
\psline(1.,-1.)(0.5,-0.5)
}

\def\OvalUpFeet {
\psline(-1.,1.)(-.5,.5)
\psline(1,1)(.5,.5)
}

\def\OvalShortFeet {
\psline(-1.,-1.)(-0.5,-0.5)
\psline(1.,-1.)(0.5,-0.5)
\psline[linewidth=.1,linestyle=dotted,dotsep=.3](-.5,-1)(.5,-1)
}

\def\DoubleCircleUp{
\pscircle(-.75,0){.5}
\pscircle(.75,0){.5}
\psline(-.5,.49)(-.3,1.0)
\psline(.5,.49)(.3,1.0)
}

\def\DoubleCircleFeet {
\psline(-1.,-1.)(-0.75,-0.5)
\psline(1.,-1.)(0.75,-0.5)
\psline[linewidth=.1,linestyle=dotted,dotsep=.3](-.5,-1)(.5,-1)
}

\pspicture(-1,0)(15,7) 
\rput(0,3.7){
\rput(0,1.5){\BSKern} 
\rput(0,1.5){\OvalShortFeet}
\rput(0,1.5){\OvalUpFeet}
\rput(1.3,1.5){{\large 2i}}
\rput(2,1.5){=}
\rput(3.5,1.5){\BSKern} 
\rput(3.5,1.5){\OvalShortFeet}
\rput(3.5,1.5){\OvalUpFeet}
\rput(4.8,1.5){{\large 1i}}
\rput(5.7,1.5){\ -  $\frac 1 2$}
\rput(7.5,1.){\BSKern} 
\rput(7.5,1.){\OvalShortTwoFeet}
\rput(7.5,2.5){\BSKern}
\rput(7.5,2.5){\large B} 
\rput(7.5,2.5){\OvalUpFeet}
\rput(8.8,1.){{\large 1i}} 
\psline(7.0,1.5)(7.0,2.0)
\psline(8.0,1.5)(8.0,2.0)
\rput(9.7,1.5){-\ $\sum$}
\rput(11.5,2.5){\BSKern}
\rput(11.5,2.5){\OvalUpFeet}
\rput(11.5,2.5){\large B}
\rput(11.5,1.){\DoubleCircleUp}
\rput(11.5,1.){\DoubleCircleFeet}
}

\rput(0,0){
\rput(0,1.5){\BSKern} 
\rput(0,1.5){\OvalShortTwoFeet}
\rput(0,1.5){\OvalUpFeet}
\rput(0,1.5){{\large B}}
\rput(2,1.5){=}
\rput(3.5,1.5){\BSKern} 
\rput(3.5,1.5){\OvalShortTwoFeet}
\rput(3.5,1.5){\OvalUpFeet}
\rput(4.8,1.5){{\large 1i}}
\rput(5.7,1.5) {\ -  $\frac 1 2$}
\rput(7.5,1.){\BSKern} 
\rput(7.5,1.){\OvalShortTwoFeet}
\rput(7.5,2.5){\BSKern}
\rput(7.5,2.5){\large B} 
\rput(7.5,2.5){\OvalUpFeet}
\rput(8.8,1.){{\large 1i}}
\psline(7.0,1.5)(7.0,2.0)
\psline(8.0,1.5)(8.0,2.0)
}

\endpspicture

\noindent The 1-particle irreducible graphs are obtained by subtracting Born terms.
%
\subsection{Dynamical equations}
for $\phi^3$-theory; can be generalized to $\phi^4$-theory by considering fields $\phi$ and
$:\phi^2$:
\psset{unit=1cm}
\psset{dotsize=0.2 0}
\psset{linewidth=.05}

\def\Ellipse{
\psellipse(0,0)(1.4,0.7)
}

\def\LineUp {
\psline(0,0)(0,1.5)
}
\def\Circle {
\pscircle(0,0){.7}
}
\def\OvalShort { 
\psline(-.5,-.5)(.5, -.5)
\psline(-.5,.5)(.5,.5)
\psarc(-.5,0){.5}{90}{270}
\psarc(.5,0){.5}{270}{90}
}
\def\OvalLong { 
\psline(-.75,-.5)(.75, -.5)
\psline(-.75,.5)(.75,.5)
\psarc(-.75,0){.5}{90}{270}
\psarc(.75,0){.5}{270}{90}
}
\def\BSKern {
\psline(-.5,-.5)(.5, -.5)
\psline(-.5,.5)(.5,.5)
\psarc(-.5,0){.5}{90}{270}
\psarc(.5,0){.5}{270}{90}
\psline[linestyle=dotted](-1.,0)(1.,0)
}
\def\CircleUp{
\pscircle(0,0){.5}
\psline(0,.5)(0,1.5)
}
\def\CircleUpAmp {
\pscircle(0,0){.5}
\psline(0,.5)(0,.7)
}
\def\CircleFeet{
\psline(-1.,-1.)(-0.3,-0.3)
\psline(1.,-1.)(0.3,-0.3)
\psline[linewidth=.1,linestyle=dotted,dotsep=.3](-.5,-1)(.5,-1)
}
\def\OvalShortFeet {
\psline(-1.,-1.)(-0.5,-0.5)
\psline(1.,-1.)(0.5,-0.5)
\psline[linewidth=.1,linestyle=dotted,dotsep=.3](-.5,-1)(.5,-1)
}
\def\DoubleCircleFeet {
\psline(-1.,-1.)(-0.75,-0.5)
\psline(1.,-1.)(0.75,-0.5)
\psline[linewidth=.1,linestyle=dotted,dotsep=.3](-.5,-1)(.5,-1)
}

\def\OvalShortInt{
\psline(-.5,-.5)(.5, -.5)
\psline(-.5,.5)(.5,.5)
\psarc(-.5,0){.5}{90}{270}
\psarc(.5,0){.5}{270}{90}
\psline(-.5,.5)(0,1.5)
\psline(.5,.5)(0,1.5)
}
\def\OvalShortIPIInt{
\psline(-.5,-.5)(.5, -.5)
\psline(-.5,.5)(.5,.5)
\psarc(-.5,0){.5}{90}{270}
\psarc(.5,0){.5}{270}{90}
\psline(-.5,.5)(0,1.5)
\psline(.5,.5)(0,1.5)
\psline[linestyle=dotted](-1.,0)(1.,0) 
}

\def\OvalShortIPICircle {
\psline(-.5,-.5)(.5, -.5)
\psline(-.5,.5)(.5,.5)
\psarc(-.5,0){.5}{90}{270}
\psarc(.5,0){.5}{270}{90}
\pscircle(0,1.5){.5}
\psline(-.5,.5)(-.2,1.1)
\psline(.5,.5)(.2,1.1)
\psline[linestyle=dotted](-1.,0)(1.,0)
\psline(0,2.0)(0,2.2) 
}

\def\DoubleCircleInt{
\pscircle(-.75,0){.5}
\pscircle(.75,0){.5}
\psline(-.5,.5)(0,1.5)
\psline(.5,.5)(0,1.5)
}

\def\DoubleCircleCircle{
\pscircle(-.75,0){.5}
\pscircle(.75,0){.5}
\psline(-.5,.5)(-.2,1.1)
\psline(.5,.5)(.2,1.1)

\pscircle(0,1.5){.5}
\psline(0,2.0)(0,2.2)
}

\def\DoubleCircle{
\pscircle(-.75,0){.5}
\pscircle(.75,0){.5}
}

\pspicture(-1,0)(13,3.8)
\rput(0,1.5){\CircleUpAmp}
\rput(0,1.5){\CircleFeet} 
\rput(-1,.1){$x_1$} 
\rput(1,.1){$x_n$} 
\rput(1.5,1.5){=}
\rput(2.2,1.5){$\quad \frac 1 2 $}
\rput(4.5,1.5){\OvalShortIPICircle}
\rput(5.8,1.5){$ 2i $}
\rput(4.5,1.5){\OvalShortFeet}
\rput(7,1.5){$\quad + \sum$}
\rput(9.5,1.5){\DoubleCircleCircle}
\rput(9.5,1.5){\DoubleCircleFeet}

\endpspicture

 The  equations for $n\geq3$ can be solved iteratively by skeleton graph 
expansions, producing also skeleton graph expansions for the Bethe Salpeter
 kernel. After that it remains to solve the renormalized Schwinger Dyson 
equations for 3-vertex and propagator:
\psset{unit=1cm}
\psset{dotsize=0.2 0}
\psset{linewidth=.05}
\def\BSKern {
\psline(-.5,-.5)(.5, -.5)
\psline(-.5,.5)(.5,.5)
\psarc(-.5,0){.5}{90}{270}
\psarc(.5,0){.5}{270}{90}
\psline[linestyle=dotted](-1.,0)(1.,0)
}

\def\Ellipse{
\psellipse(0,0)(1.4,0.7)
}

\def\LineUp {
\psline(0,0)(0,1.5)
}
\def\Circle {
\pscircle(0,0){.7}
}
\def\OvalShort { 
\psline(-.5,-.5)(.5, -.5)
\psline(-.5,.5)(.5,.5)
\psarc(-.5,0){.5}{90}{270}
\psarc(.5,0){.5}{270}{90}
}
\def\OvalLong { 
\psline(-.75,-.5)(.75, -.5)
\psline(-.75,.5)(.75,.5)
\psarc(-.75,0){.5}{90}{270}
\psarc(.75,0){.5}{270}{90}
}
\def\BSKern {
\psline(-.5,-.5)(.5, -.5)
\psline(-.5,.5)(.5,.5)
\psarc(-.5,0){.5}{90}{270}
\psarc(.5,0){.5}{270}{90}
\psline[linestyle=dotted](-1.,0)(1.,0)
}
\def\CircleUp{
\pscircle(0,0){.5}
\psline(0,.5)(0,1.5)
\rput(1,0.5){${}^\prime$} 
}
\def\CircleUpAmp {
\pscircle(0,0){.5}
\psline(0,.5)(0,.7)
\rput(1,0.5){${}^\prime$} 
}
\def\CircleFeet{
\psline(-1.,-1.)(-0.3,-0.3)
\psline(1.,-1.)(0.3,-0.3)
\psline[linewidth=.1,linestyle=dotted,dotsep=.3](-.5,-1)(.5,-1)
}

\def\CircleTwoFeet{
\psline(-1.,-1.)(-0.3,-0.3)
\psline(1.,-1.)(0.3,-0.3)
}
\def\OvalShortFeet {
\psline(-1.,-1.)(-0.5,-0.5)
\psline(1.,-1.)(0.5,-0.5)
\rput(0,0){{\large B}}
}
\def\DoubleCircleFeet {
\psline(-1.,-1.)(-0.75,-0.5)
\psline(1.,-1.)(0.75,-0.5)
\psline[linewidth=.1,linestyle=dotted,dotsep=.3](-.5,-1)(.5,-1)
}

\def\OvalShortInt{
\psline(-.5,-.5)(.5, -.5)
\psline(-.5,.5)(.5,.5)
\psarc(-.5,0){.5}{90}{270}
\psarc(.5,0){.5}{270}{90}
\psline(-.5,.5)(0,1.5)
\psline(.5,.5)(0,1.5)
}
\def\OvalShortIPIInt{
\psline(-.5,-.5)(.5, -.5)
\psline(-.5,.5)(.5,.5)
\psarc(-.5,0){.5}{90}{270}
\psarc(.5,0){.5}{270}{90}
\psline(-.5,.5)(0,1.5)
\psline(.5,.5)(0,1.5)
\psline[linestyle=dotted](-1.,0)(1.,0) 
}

\def\OvalShortIPICircle {
\psline(-.5,-.5)(.5, -.5)
\psline(-.5,.5)(.5,.5)
\psarc(-.5,0){.5}{90}{270}
\psarc(.5,0){.5}{270}{90}
\pscircle(0,1.5){.5}
\psline(-.5,.5)(-.2,1.1)
\psline(.5,.5)(.2,1.1)
\psline[linestyle=dotted](-1.,0)(1.,0)
\psline(0,2.0)(0,2.2) 
}

\def\DoubleCircleInt{
\pscircle(-.75,0){.5}
\pscircle(.75,0){.5}
\psline(-.5,.5)(0,1.5)
\psline(.5,.5)(0,1.5)
}

\def\DoubleCircleCircle{
\pscircle(-.75,0){.5}
\pscircle(.75,0){.5}
\psline(-.5,.5)(-.2,1.1)
\psline(.5,.5)(.2,1.1)

\pscircle(0,1.5){.5}
\psline(0,2.0)(0,2.2)
}

\def\DoubleCircle{
\pscircle(-.75,0){.5}
\pscircle(.75,0){.5}
}

\pspicture(-1.5,1)(13,3.7) 
\rput(0,1.5){\CircleUpAmp}
\rput(0,1.5){\CircleTwoFeet} 
\rput(1.5,1.5){=}
\rput(2.2,1.5){$\quad \frac 1 2 $}
\rput(4.5,1.5){\OvalShortIPICircle}
\rput(5.5,3.5){${}^\prime$}
\rput(4.5,1.5){\BSKern}
\rput(4.5,1.5){\OvalShortFeet}

\endpspicture

\psset{unit=1cm}
\psset{dotsize=0.2 0}
\psset{linewidth=.05}

\def\Ellipse{
\psellipse(0,0)(1.4,0.7)
}

\def\LineUp {
\psline(0,0)(0,1.5)
}
\def\Circle {
\pscircle(0,0){.7}
}
\def\OvalShort { 
\psline(-.5,-.5)(.5, -.5)
\psline(-.5,.5)(.5,.5)
\psarc(-.5,0){.5}{90}{270}
\psarc(.5,0){.5}{270}{90}
}
\def\OvalLong { 
\psline(-.75,-.5)(.75, -.5)
\psline(-.75,.5)(.75,.5)
\psarc(-.75,0){.5}{90}{270}
\psarc(.75,0){.5}{270}{90}
}
\def\BSKern {
\psline(-.5,-.5)(.5, -.5)
\psline(-.5,.5)(.5,.5)
\psarc(-.5,0){.5}{90}{270}
\psarc(.5,0){.5}{270}{90}
\psline[linestyle=dotted](-1.,0)(1.,0)
}

\def\BSKernArrow {
\psline[linewidth=3pt]{->}(.45,-.5)(.45,.5) 
\psline(-.5,-.5)(.5, -.5)
\psline(-.5,.5)(.5,.5)
\psarc(-.5,0){.5}{90}{270}
\psarc(.5,0){.5}{270}{90}
\psline[linestyle=dotted](-1.,0)(1.,0)
}

\def\CircleUp{
\pscircle(0,0){.5}
\psline(0,.5)(0,1.5)
}
\def\CircleUpAmp {
\pscircle(0,0){.5}
\psline(0,.5)(0,.7)
}
\def\CircleFeet{
\psline(-1.,-1.)(-0.3,-0.3)
\psline(1.,-1.)(0.3,-0.3)
\psline[linewidth=.1,linestyle=dotted,dotsep=.3](-.5,-1)(.5,-1)
}
\def\OvalShortFeet {
\psline(-1.,-1.)(-0.5,-0.5)
\psline(1.,-1.)(0.5,-0.5)
\psline[linewidth=.1,linestyle=dotted,dotsep=.3](-.5,-1)(.5,-1)
}
\def\DoubleCircleFeet {
\psline(-1.,-1.)(-0.75,-0.5)
\psline(1.,-1.)(0.75,-0.5)
\psline[linewidth=.1,linestyle=dotted,dotsep=.3](-.5,-1)(.5,-1)
}

\def\OvalShortInt{
\psline(-.5,-.5)(.5, -.5)
\psline(-.5,.5)(.5,.5)
\psarc(-.5,0){.5}{90}{270}
\psarc(.5,0){.5}{270}{90}
\psline(-.5,.5)(0,1.5)
\psline(.5,.5)(0,1.5)
}
\def\OvalShortIPIInt{
\psline(-.5,-.5)(.5, -.5)
\psline(-.5,.5)(.5,.5)
\psarc(-.5,0){.5}{90}{270}
\psarc(.5,0){.5}{270}{90}
\psline(-.5,.5)(0,1.5)
\psline(.5,.5)(0,1.5)
\psline[linestyle=dotted](-1.,0)(1.,0) 
}

\def\OvalShortIPICircle {
\psline(-.5,-.5)(.5, -.5)
\psline(-.5,.5)(.5,.5)
\psarc(-.5,0){.5}{90}{270}
\psarc(.5,0){.5}{270}{90}
\pscircle(0,1.5){.5}
\psline(-.5,.5)(-.2,1.1)
\psline(.5,.5)(.2,1.1)
\psline[linestyle=dotted](-1.,0)(1.,0)
\psline(0,2.0)(0,2.2) 
}

\def\DoubleCircleInt{
\pscircle(-.75,0){.5}
\pscircle(.75,0){.5}
\psline(-.5,.5)(0,1.5)
\psline(.5,.5)(0,1.5)
}

\def\DoubleCircleCircle{
\pscircle(-.75,0){.5}
\pscircle(.75,0){.5}
\psline(-.5,.5)(-.2,1.1)
\psline(.5,.5)(.2,1.1)

\pscircle(0,1.5){.5}
\psline(0,2.0)(0,2.2)
}

\def\DoubleCircle{
\pscircle(-.75,0){.5}
\pscircle(.75,0){.5}
}

\def\OvalShortTwoFeet {
\psline(-1.,-1.)(-0.5,-0.5)
\psline(1.,-1.)(0.5,-0.5)
}

\def\OvalShortTwoShortFeet {
\psline(-0.6,-0.6)(-0.5,-0.5)
\psline(.6,-0.6)(0.5,-0.5)
}

\def\OvalShortOneShortFoot {
\psline(0,-0.6)(0,-0.5)
}

\def\OvalUpFeet {
\psline(-1.,1.)(-.5,.5)
\psline(1,1)(.5,.5)
}

\def\OvalUpShortFeet {
\psline(-.6,.6)(-.5,.5)
\psline(.6,.6)(.5,.5)
}
 \def\OvalUpShortFoot {
\psline(0,.6)(0,.5)
}

\pspicture(-1,-0.5)(12,5)
\rput(1.2,0){
 \psline[linewidth=3pt,linecolor=black]{->}(-1.4,3.3)(-.7,3.3)
 \rput(1.5,3.3){= \ $x^\mu - y^\mu$}
 \rput(-.7,2.75){$y$}
 \rput(-1.4,2.8){$x$}
}
 \rput(0.,1.5){0}
 \rput(1.5,1.5){ = \ \ $\frac 1 2 $}
 \rput(-3.5,-0.25){
\rput(7.5,1.){\OvalShort} 
\rput(7.5,1.){\OvalShortOneShortFoot}
\rput(7.5,2.5){\OvalShort}
\rput(7.5,2.5){\OvalUpShortFoot}
%
\psline(7.0,1.5)(7.0,2.0)
\psline[linewidth=3pt]{->}(8.0,1.5)(8.0,2.0)
}
\rput(5.7,1.5){+ \ $\frac 1 4$}
\rput(0,-1.){
\rput(7.5,1.){\OvalShort} 
\rput(7.5,1.){\OvalShortOneShortFoot}
\rput(7.5,2.5){\BSKernArrow}
\rput(7.5,2.5){\large B}

\psline(7.0,1.5)(7.0,2.0)
\psline(8.0,1.5)(8.0,2.0)
\rput(7.5,4){\OvalShort} 
\rput(7.5,4){\OvalUpShortFoot} 
\psline(7.0,3.0)(7.0,3.5)
\psline(8.0,3.0)(8.0,3.5)

}
\endpspicture

The equation for the inverse propagator is actually equivalent to an equation for the 3-point function of the stress enegy tensor \cite{MackSymanzik}.
\subsection{Skeleton graph expansions}
The equations for $n\geq 4$ point functions can be solved iteratively, producing skeleton graph expansions for Green functions and also for the Bethe Salpeter kernel \cite{MackSymanzik}:
\psset{unit=.8cm}
\psset{dotsize=0.2 0}
\psset{linewidth=.05}
\def\BSKern{
\psline(-.5,-.5)(.5, -.5)
\psline(-.5,.5)(.5,.5)
\psarc(-.5,0){.5}{90}{270}
\psarc(.5,0){.5}{270}{90}
\psline[linestyle=dotted](-1.,0)(1.,0)
\rput(0,0.05){$B$}
}

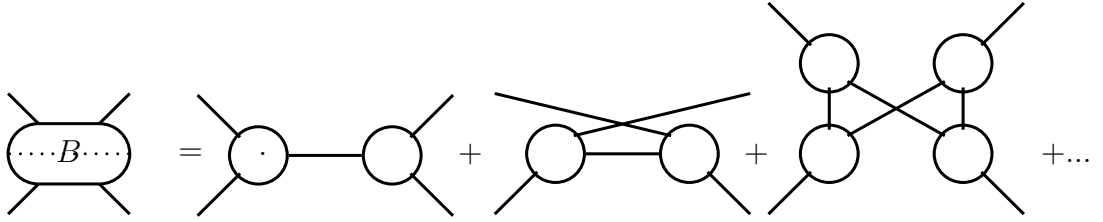
\def\BSKernWithFeet{
  \rput(0,0){
  \BSKern
  \OvalUpFeet 
  \OvalShortTwoFeet 
 }
}
\def\OvalUpFeet {
\psline(-1.,1.)(-.5,.5)
\psline(1,1)(.5,.5)
}
\def\OvalShortTwoFeet {
\psline(-1.,-1.)(-0.5,-0.5)
\psline(1.,-1.)(0.5,-0.5)
}
\def\PartialWave{
\rput{-90}(0,0){
\CGKern
\CircleFeet
 }
\rput{90}(2.2,0){
\CGKern
\CircleFeet
 }
}
\def\PartialWaveFeetDown{
\rput(0,0){\CGKern}
\rput(2.2,0){\CGKern}
\psline(.5,0)(1.7,0)
}
\def\CGKern{
\pscircle(0,0){.5}
\rput(0,0.5){\zigzagTwo}
}
\def\CircleFeet{
\psline(-1.,-1.)(-0.3,-0.3)
\psline(1.,-1.)(0.3,-0.3)
}
\def\zigzagTwo{
\psline(0,0)(0.1,0.1)(-0.1,0.3)(0.1,0.5)(0,0.6)
}
\def\Vertex{
\pscircle(0,0){.5}
}
\def\BornTerm{
\rput(0,0){\Vertex}
\rput(2.2,0){\Vertex}
\psline(.5,0)(1.7,0)
\psline(-.3,-.3)(-1.,-1.)
\psline(2.5,-.3)(3.2,-1.)
\psline(-.3,.3)(-1.,1)   
\psline(2.5,.3)(3.2,1.). 
}
\def\BornTermDownFeet
{
\rput(0,0){\Vertex}
\rput(2.2,0){\Vertex}
\psline(.5,0)(1.7,0)
\psline(-.3,-.3)(-1.,-1.)
\psline(2.5,-.3)(3.2,-1.)
}
\def\BornTermCrossed{
\BornTermDownFeet
\psline(0.3,0.3)(3.2,1.)
\psline(1.9,0.3)(-1.,1)
}
\def\SecondGraph{
\rput(0,0){\Vertex}
\rput(2.2,0){\Vertex}
\rput(0,1.5){\Vertex}
\rput(2.2,1.5){\Vertex}
\psline(0,0.4)(0,1.1) 

\psline(2.2,.4)(2.2,1.1)
\psline(0.3,0.3)(1.9,1.2)
\psline(1.9,0.3)(0.3,1.2)
\psline(-.3,-.3)(-1.,-1.)
\psline(2.5,-.3)(3.2,-1.)
\psline(-.3,1.8)(-1.,2.5)
\psline(2.5,1.8)(3.2,2.5)
}

\pspicture(-1.2,-1)(11,3.5)
\rput(0,1){
\rput(0,.0){\BSKernWithFeet}
\rput(2,0){$=$}
\rput(3.3,0){\BornTerm} 
\rput(6.6,0){$+$}
\rput(8.0,0){\BornTermCrossed}
\rput(11.3,0){$+$}
\rput(12.5,0){\SecondGraph}
\rput(16.4,0){$+ ...$}
}
\endpspicture

In $\phi^4$-theory, Born terms involve $:\phi^2:$ propagators
\subsection{$\epsilon$-expansions}
In $D=6+\epsilon$ dimensions for $\varphi^3$-theory (and in $D=4-\epsilon$ dimensions for $\phi^4$-theory), the dressed vertices are of order $\epsilon$. Therefore only a finite number of skeleton graphs contribute to any order in $\epsilon$. Inserting into the equations for dressed vertex and propagator, these can be solved by power series expansion in $\epsilon$. For $\phi^3$-theory in 
$6+\epsilon$ dimensions one finds to leading orders \cite{Mack3}: The 
fundamental field $\phi$ has dimension
$$ d= \frac{D}{2} -1 + \Delta,\qquad \qquad \Delta= \frac {1}{18}\epsilon + ...$$ 
and a trajectory of traceless symmetric tensor fields of even rank $s\geq 2$ with  dimension $d_s= D-2 + s +\sigma_s$,
$$ \half \sigma_s= \left[ \frac {1}{18} - \frac{2}{3(s+2)(s+1)}\right]
 \epsilon + ..
$$

\section{\large The relation between reprentations and harmonic analysis for covering groups of $SO(D,2)$ and $SO(D+1,1)$}
 Basic: Elementary representations $\chi=[l,\delta]$
{\bf are representations of the complex Lie algebra}.
 Partial wave amplitudes $g(\chi)$ depend on $\chi$. Their use is a crucial feature of the group theoretical approach \cite{Mack1,Mack2,Mack3,DobrevPetkovaPetrovaTodorov,DobrevMackPetkovaPetrovaTodorov}.
 The (partial) equivalence of representations $\chi=[l,\delta]$ and $\tilde{\chi}=[\tilde{l}, D-\delta]$ ($\tilde{l}=l$ for completely symmetric tensor 
representations) is the basis of the shadow operator formalism introduced by Ferrara et al. \cite{FerraraGattoGrillo}
\\
\subsection{Euklidean partial wave expansion}
for the 4-point function $<\phi(x_4)...\phi(x_1)>$. 
\begin{figure}[h]
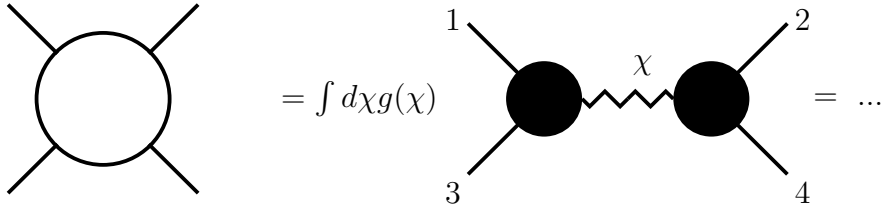

\psset{unit=1cm}
\psset{dotsize=0.2 0}
\psset{linewidth=.05}
\def\CircleFeet{
\psline(-1.,-1.)(-0.3,-0.3)
\psline(1.,-1.)(0.3,-0.3)
}
\def\OvalShort { 
\psline(-.5,-.5)(.5, -.5)
\psline(-.5,.5)(.5,.5)
\psarc(-.5,0){.5}{90}{270}
\psarc(.5,0){.5}{270}{90}
}
\def\OvalLong { 
\psline(-.75,-.5)(.75, -.5)
\psline(-.75,.5)(.75,.5)
\psarc(-.75,0){.5}{90}{270}
\psarc(.75,0){.5}{270}{90}
}
\def\BSKern {
\psline(-.5,-.5)(.5, -.5)
\psline(-.5,.5)(.5,.5)
\psarc(-.5,0){.5}{90}{270}
\psarc(.5,0){.5}{270}{90}
\psline[linestyle=dotted](-1.,0)(1.,0)
}

\def\CircleUp{
\pscircle(0,0){.5}
\psline(0,.5)(0,1.5)
}
\def\OvalShortTwoFeet {
\psline(-1.,-1.)(-0.5,-0.5)
\psline(1.,-1.)(0.5,-0.5)
}

\def\OvalUpFeet {
\psline(-1.,1.)(-.5,.5)
\psline(1,1)(.5,.5)
}

\def\OvalShorFteet {
\psline(-1.,-1.)(-0.5,-0.5)
\psline(1.,-1.)(0.5,-0.5)
}
\def\zigzagOne{
\psline(0,0)(0.1,0.1)(-0.1,0.3)(0,0.4)
}
\def\zigzagTwo{
\psline(0,0)(0.1,0.1)(-0.1,0.3)(0.1,0.5)(0,0.6)
}
\def\CGKern{
\pscircle*(0,0){.5}
\rput(0,0.5){\zigzagTwo}
}

\def\PartialWave{
\rput{-90}(0,0){
\CGKern
\CircleFeet
 }
\rput{90}(2.2,0){
\CGKern
\CircleFeet
 }
}
\def\PartialWaveUp{
\rput{90}(0,0){\PartialWave}
}
\def\FourPointFct{
\pscircle(0,0){0.9}
\psline(-1.25,-1.25)(-0.60,-0.60)
\psline(0.60,0.60)(1.25,1.25)
%
\psline(1.25,-1.25)(0.60,-0.60)
\psline(-0.60,0.60)(-1.25,1.25)
}

\def\FourPointMellin{
\rput(-2.9,0){$=\int d\delta \ M(\{ \delta_{ij}\})$}
\psline(-1.25,-1.25)(-1.25,1.25)
\psline(-1.25,-1.25)(1.25,-1.25)
\psline(-1.25,-1.25)(1.25,1.25)
\psline(1.25,1.25)(-1.25,1.25)
\psline(1.25,1.25)(1.25,-1.25)
\psline(1.25,-1.25)(-1.25,1.25)
}

\pspicture(0,0)(11,3.0)

\rput(7,0){
\rput(-2.5,1.25){ $ = \int d\chi  g(\chi) $}
\rput(-.0,0){
\rput(-1.2,0){$3$}
\rput(-1.2,2.3){$1$}
\rput(3.4,0){$4$}
\rput(3.4,2.3){$2$}
\rput(0.0,1.25){\PartialWave} 
\rput(1.3,1.7){$\chi$}
}
}
\rput(1.2,1.25){\FourPointFct}
\rput(11,1.25){$=\  ...$}
%
%
\endpspicture
\caption{Euklidean partial wave expansion for the 4-point function \newline
 $<\phi(x_4)...\phi(x_1)>$.
The dots stand for expansions in the two other channels $(12)\mapsto (34) $ and  $(13)\mapsto (24)$ of the same amplitude, implying the equality shown in section 4.2
 \label{fig:EuklideanPartialWave.tex}}
\end{figure}


This formula {\em is quite trivial, simply expressing the orthogonality} (and completeness!) {\em of the partial waves} say Simmons-Duffin, Stanford, and Witten \cite{SimmonsDuffinStanfordWitten}.

 indeed: Physical positivity implies constraints on the partial waves $g(\chi)$! 
But also
 completeness is not trivial, because presence of a Born term requires contributions from the {\em complementary} series in addition to the 
{\em principal} series.

 They can
 be included by choosing a proper path of the $c$-integration
$$\int d\chi .... = \sum_l\int_C dc \rho_l(c)...,\qquad \chi=[l,\half D + c]$$
\setlength{\unitlength}{0.65mm}
\begin{picture}(100,60)(-2,0)
\thinlines
\put(0,10){
\put(38,25){\line(1,0){4}}
\thicklines
\put(40,0){\vector(0,1){50}}
\put(32,25){\circle*{3}}
\put(30,15){$c_f$}
\put(48,25){\circle{3}}
\put(32,25){\circle{9}}
\put(36.4,24.5){\vector(0,-1){1}}
\put(48,25){\circle{9}}
\put(43.6,24.5){\vector(0,-1){1}}

\put(62,25){\circle*{3}}
\put(74,25){\circle*{3}}
\put(84,25){\circle*{3}}

\put(18,25){\circle{3}}
\put(6,25){\circle{3}}
\put(-4,25){\circle{3}}
}
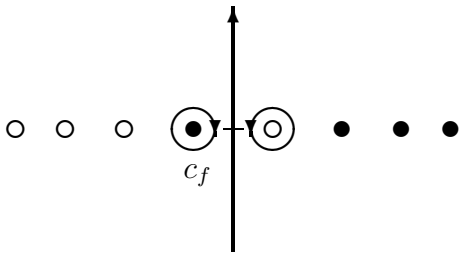
\put(0,3){a) Path $C$ of the  c-integration for $l=0$}
\end{picture}


{\bf Positivity constraints} were examined already in the 7o's. 
They restrict the position of the
 poles and fix the sign of of the residues of the partial wave amplitudes $g(\chi)$ defined by
\psset{unit=1cm}
\psset{dotsize=0.2 0}
\psset{linewidth=.05}
\def\CircleFeet{
\psline(-1.,-1.)(-0.3,-0.3)
\psline(1.,-1.)(0.3,-0.3)
}
\def\OvalShort { 
\psline(-.5,-.5)(.5, -.5)
\psline(-.5,.5)(.5,.5)
\psarc(-.5,0){.5}{90}{270}
\psarc(.5,0){.5}{270}{90}
}
\def\OvalLong { 
\psline(-.75,-.5)(.75, -.5)
\psline(-.75,.5)(.75,.5)
\psarc(-.75,0){.5}{90}{270}
\psarc(.75,0){.5}{270}{90}
}
\def\BSKern {
\psline(-.5,-.5)(.5, -.5)
\psline(-.5,.5)(.5,.5)
\psarc(-.5,0){.5}{90}{270}
\psarc(.5,0){.5}{270}{90}
\psline[linestyle=dotted](-1.,0)(1.,0)
}

\def\CircleUp{
\pscircle(0,0){.5}
\psline(0,.5)(0,1.5)
}
\def\OvalShortTwoFeet {
\psline(-1.,-1.)(-0.5,-0.5)
\psline(1.,-1.)(0.5,-0.5)
}

\def\OvalUpFeet {
\psline(-1.,1.)(-.5,.5)
\psline(1,1)(.5,.5)
}

\def\OvalShorFteet {
\psline(-1.,-1.)(-0.5,-0.5)
\psline(1.,-1.)(0.5,-0.5)
}
\def\zigzagOne{
\psline(0,0)(0.1,0.1)(-0.1,0.3)(0,0.4)
}
\def\zigzagTwo{
\psline(0,0)(0.1,0.1)(-0.1,0.3)(0.1,0.5)(0,0.6)
}
\def\CGKern{
\pscircle*(0,0){.5}
\rput(0,0.5){\zigzagTwo}
}

\def\PartialWave{
\rput{-90}(0,0){
\CGKern
\CircleFeet
 }
\rput{90}(2.2,0){
\CGKern
\CircleFeet
 }
}
\def\PartialWaveUp{
\rput{90}(0,0){\PartialWave}
}
\def\FourPointFct{
\pscircle(0,0){0.9}
\psline(-1.25,-1.25)(-0.60,-0.60)
\psline(0.60,0.60)(1.25,1.25)
%
\psline(1.25,-1.25)(0.60,-0.60)
\psline(-0.60,0.60)(-1.25,1.25)
}
\def\FourPointFctLeftOnly{
\pscircle(0,0){0.9}
\psline(-1.25,-1.25)(-0.60,-0.60)
\psline(-1.25,1.25)(-0.60,0.60)
%
}

\def\FourPointMellin{
\rput(-2.9,0){$=\int d\delta \ M(\{ \delta_{ij}\})$}
\psline(-1.25,-1.25)(-1.25,1.25)
\psline(-1.25,-1.25)(1.25,-1.25)
\psline(-1.25,-1.25)(1.25,1.25)
\psline(1.25,1.25)(-1.25,1.25)
\psline(1.25,1.25)(1.25,-1.25)
\psline(1.25,-1.25)(-1.25,1.25)
}
\def\rhspw{
  \FourPointFctLeftOnly
  \rput(2,0){\rotatebox{-90}{\CGKern}}
  \psline(.6,.6)(2.0,0)
  \psline(.6,-.6)(2.0,0)
  \rput(3.5,0){$\chi$}
}


\pspicture(0,0)(11,2.5)
\rput(0.5,1){
 \rput(0.1,0){$g(\chi)$}
 \rput(1.3,0){\rotatebox{-90}{
   \CGKern
   \CircleFeet
  } 
 }
 \rput(2.8,0){$\chi =$}
 \rput(4.5,0){\rhspw}
}
\endpspicture


\subsection{Crossing relations}

\psset{unit=1cm}
\psset{dotsize=0.2 0}
\psset{linewidth=.05}
\def\CircleFeet{
\psline(-1.,-1.)(-0.3,-0.3)
\psline(1.,-1.)(0.3,-0.3)
}
\def\OvalShort { 
\psline(-.5,-.5)(.5, -.5)
\psline(-.5,.5)(.5,.5)
\psarc(-.5,0){.5}{90}{270}
\psarc(.5,0){.5}{270}{90}
}
\def\OvalLong { 
\psline(-.75,-.5)(.75, -.5)
\psline(-.75,.5)(.75,.5)
\psarc(-.75,0){.5}{90}{270}
\psarc(.75,0){.5}{270}{90}
}
\def\BSKern {
\psline(-.5,-.5)(.5, -.5)
\psline(-.5,.5)(.5,.5)
\psarc(-.5,0){.5}{90}{270}
\psarc(.5,0){.5}{270}{90}
\psline[linestyle=dotted](-1.,0)(1.,0)
}

\def\CircleUp{
\pscircle(0,0){.5}
\psline(0,.5)(0,1.5)
}
\def\OvalShortTwoFeet {
\psline(-1.,-1.)(-0.5,-0.5)
\psline(1.,-1.)(0.5,-0.5)
}

\def\OvalUpFeet {
\psline(-1.,1.)(-.5,.5)
\psline(1,1)(.5,.5)
}

\def\OvalShorFteet {
\psline(-1.,-1.)(-0.5,-0.5)
\psline(1.,-1.)(0.5,-0.5)
}
\def\zigzagOne{
\psline(0,0)(0.1,0.1)(-0.1,0.3)(0,0.4)
}
\def\zigzagTwo{
\psline(0,0)(0.1,0.1)(-0.1,0.3)(0.1,0.5)(0,0.6)
}
\def\CGKern{
\pscircle*(0,0){.5}
\rput(0,0.5){\zigzagTwo}
}

\def\PartialWave{
\rput{-90}(0,0){
\CGKern
\CircleFeet
 }
\rput{90}(2.2,0){
\CGKern
\CircleFeet
 }
}

\pspicture(-1,0)(11,3.5)
\rput(1,0.5){
\rput(0.5 ,1.25){ $ \int d\chi \quad g(\chi) $}
\rput(3,0){
\rput(-1.2,0){$3$}
\rput(-1.2,2.3){$1$}
\rput(3.4,0){$4$}
\rput(3.4,2.3){$2$}
\rput(0,1.25){\PartialWave}
\rput(1.3,1.7){$\chi$}
}
\rput(2,0){
\rput(6,1.25){$\quad =\int d\chi^\prime g(\chi^\prime)$} 
\rput{90}(9,0.5){\PartialWave}
\rput(9.75,1.75){$\chi^\prime$}
\rput(7.8,3.8){$1$}
\rput(10.2,3.8){$2$}
\rput(10.2,-0.4){$4$}
\rput(7.8,-0.4){$3$}
 }
}
\endpspicture


%
%

\noindent The crossing relation is a linear relation for the partial wave, 
of the form $$g(\chi)= \int d\chi^\prime C(\chi^\prime, \chi)g(\chi^\prime).$$ Using orhogonality of the CG-kernels, the crossing kernel 
$C(\chi,\chi^\prime)$ is defined by
\psset{unit=1cm}
\psset{dotsize=0.2 0}
\psset{linewidth=.05}
\def\CircleLeftFoot{
 \psline(-1.,-1.)(-0.3,-0.3)
}
\def\CircleLongLeftFoot{
 \psline(-1.,-1.)(1.,1.)
}
\def\CircleRightFoot{
\psline(1.,-1.)(0.3,-0.3)
}
\def\CircleLongRightFoot{
\psline(1.,-1.)(-1.,1.)
}
\def\CircleFeet{
\psline(-1.,-1.)(-0.3,-0.3)
\psline(1.,-1.)(0.3,-0.3)
}
\def\OvalShort { 
\psline(-.5,-.5)(.5, -.5)
\psline(-.5,.5)(.5,.5)
\psarc(-.5,0){.5}{90}{270}
\psarc(.5,0){.5}{270}{90}
}
\def\OvalLong { 
\psline(-.75,-.5)(.75, -.5)
\psline(-.75,.5)(.75,.5)
\psarc(-.75,0){.5}{90}{270}
\psarc(.75,0){.5}{270}{90}
}
\def\BSKern {
\psline(-.5,-.5)(.5, -.5)
\psline(-.5,.5)(.5,.5)
\psarc(-.5,0){.5}{90}{270}
\psarc(.5,0){.5}{270}{90}
\psline[linestyle=dotted](-1.,0)(1.,0)
}

\def\CircleUp{
\pscircle(0,0){.5}
\psline(0,.5)(0,1.5)
}
\def\OvalShortTwoFeet {
\psline(-1.,-1.)(-0.5,-0.5)
\psline(1.,-1.)(0.5,-0.5)
}

\def\OvalUpFeet {
\psline(-1.,1.)(-.5,.5)
\psline(1,1)(.5,.5)
}

\def\OvalShorFteet {
\psline(-1.,-1.)(-0.5,-0.5)
\psline(1.,-1.)(0.5,-0.5)
}
\def\zigzagOne{
\psline(0,0)(0.1,0.1)(-0.1,0.3)(0,0.4)
}
\def\zigzagTwo{
\psline(0,0)(0.1,0.1)(-0.1,0.3)(0.1,0.5)(0,0.6)
}
\def\CGKern{
\pscircle*(0,0){.5}
\rput(0,0.5){\zigzagTwo}
}

\def\PartialWave{
\rput{-90}(0,0){
\CGKern
\CircleFeet
 }
\rput{90}(2.2,0){
\CGKern
\CircleFeet
 }
}
\def\PartialWaveFeetDown{
\rput{-90}(0,0){
\CGKern
  }
 \CircleLongLeftFoot
\rput{90}(2.2,0){
 \CGKern
  }
 \rput(2.0,0){\CircleLongRightFoot}
}
\def\PartialWaveUp{
\rput{90}(0,0){\PartialWave}
}
\def\FourPointFct{
\pscircle(0,0){0.9}
\psline(-1.25,-1.25)(-0.60,-0.60)
\psline(0.60,0.60)(1.25,1.25)
\psline(1.25,-1.25)(0.60,-0.60)
\psline(-0.60,0.60)(-1.25,1.25)
}

\def\FourPointMellin{
\rput(-2.9,0){$=\int d\delta \ M(\{ \delta_{ij}\})$}
\psline(-1.25,-1.25)(-1.25,1.25)
\psline(-1.25,-1.25)(1.25,-1.25)
\psline(-1.25,-1.25)(1.25,1.25)
\psline(1.25,1.25)(-1.25,1.25)
\psline(1.25,1.25)(1.25,-1.25)
\psline(1.25,-1.25)(-1.25,1.25)
}

\pspicture(-1,0)(5,3.4)

\rput(7,0){
 \rput(-.5,0){
 \rput(0,1.25){\PartialWaveFeetDown} 
 \rput(1.2,0.7){$\chi$}
 \rput(1.6,3.4){$\chi^\prime$}
 }
\rput(0.3,2.5){\rotatebox{90}(\CGKern} 
}
\rput(0.5,2.0){\CGKern
\CircleFeet
\rput(0.5,1.2){$\chi^\prime$}
}
\rput(3.5,2.){$C(\chi,\chi^\prime)\ = \ \frac {1}{2}$}
\endpspicture

%

\subsection{\large Derivation of Operator product expansions 
 from Euklidean partial wave expansions}
\subsubsection{Split of the partial waves}
Operator product expansions (OPE) have the structure \\
$$ \phi^1(-\half x_1)\phi^2(\half x_1)=\sum_n C_n(x_1)O^n(0)$$
where $O^n$ are fields or derivatives of fields.
The derivatives of each nonderivative field can be summed up. The resulting 
terms are known as {\bf confornal blocks}.
 To obtain this expansion, \\ 
{\bf the Clebsch Gordan coefficients need to be split.}


\psset{unit=1cm}
\psset{dotsize=0.2 0}
\psset{linewidth=.05}
\def\CircleFeet{
\psline(-1.,-1.)(-0.3,-0.3)
\psline(1.,-1.)(0.3,-0.3)
}
\def\OvalShort { 
\psline(-.5,-.5)(.5, -.5)
\psline(-.5,.5)(.5,.5)
\psarc(-.5,0){.5}{90}{270}
\psarc(.5,0){.5}{270}{90}
}
\def\OvalLong { 
\psline(-.75,-.5)(.75, -.5)
\psline(-.75,.5)(.75,.5)
\psarc(-.75,0){.5}{90}{270}
\psarc(.75,0){.5}{270}{90}
}
\def\BSKern {
\psline(-.5,-.5)(.5, -.5)
\psline(-.5,.5)(.5,.5)
\psarc(-.5,0){.5}{90}{270}
\psarc(.5,0){.5}{270}{90}
\psline[linestyle=dotted](-1.,0)(1.,0)
}

\def\CircleUp{
\pscircle(0,0){.5}
\psline(0,.5)(0,1.5)
}
\def\OvalShortTwoFeet {
\psline(-1.,-1.)(-0.5,-0.5)
\psline(1.,-1.)(0.5,-0.5)
}

\def\OvalUpFeet {
\psline(-1.,1.)(-.5,.5)
\psline(1,1)(.5,.5)
}

\def\OvalShorFteet {
\psline(-1.,-1.)(-0.5,-0.5)
\psline(1.,-1.)(0.5,-0.5)
}
\def\zigzagOne{
\psline(0,0)(0.1,0.1)(-0.1,0.3)(0,0.4)
}
\def\zigzagTwo{
\psline(0,0)(0.1,0.1)(-0.1,0.3)(0.1,0.5)(0,0.6)
}

\def\zigzagTwoHor{
\psline(0,0)(0.1,0.1)(0.3,-0.1)(0.5,0.1)(0.6,0)
}
\def\CGKern{
\pscircle*(0,0){.5}
\rput(0,0.5){\zigzagTwo}
}

\def\PartialWave{
\rput{-90}(0,0){
\CGKern
\CircleFeet
 }
\rput{90}(2.2,0){
\CGKern
\CircleFeet
 }
}
\def\CGhalf{
\psarc*(0,0){.5}{180}{0}
\psarc(0,0){.5}{0}{180}
}
\def\CGhalfHor{
\psarc*(0,0){.5}{90}{-90}
\psarc(0,0){.5}{-90}{90}
}
\def\QKern{
\psarc*(0,0){.5}{90}{-90}
\psarc(0,0){.5}{-90}{90}
\psline(-1.,-1)(-.3,-.3)
\psline(-1,1)(-.3,.3)
\rput(0.5,0){\zigzagTwoHor}
\rput(0.7,0.5){$\chi$}
}

\def\QKerntilde{
\psarc*(0,0){.5}{90}{-90}
\psarc(0,0){.5}{-90}{90}
\psline(-1.,-1)(-.3,-.3)
\psline(-1,1)(-.3,.3)
\rput(0.5,0){\zigzagTwoHor}
\rput(0.7,0.5){$\tilde{\chi}$}
\rput(1.2,0.0){\zigzaglongHor}
}
\def\zigzaglongHor{
\psline(0,0)(0.1,0.1)(0.3,-0.1)(0.5,0.1)(0.7,-0.1)(0.8,0)
}
\def\SplitCGrhs{
 \QKern 
 \rput(1.5,0){$+$}
 \rput(2.7,0){\QKerntilde}
}
\def\SplitCGlhs{
  \rotatebox{-90}{\CGKern}
  \psline(-1.,-1)(-.3,-.3)
  \psline(-1,1)(-.3,.3)
}
\pspicture(0,0)(11,2.)
\rput(1,1){
 \rput(0,0){\SplitCGlhs}
 \rput(1.7,0.0){$=$}
 \rput(3.0,0){\SplitCGrhs}
}
\endpspicture

This split exploits the equivalence of representations $\chi=[l,h+c]$ and 
$\tilde{\chi}=[l,h-c]$ $(h=\half D)$ of the Euklidean conformal group to split
 the three point function into Clebsch-Gordan kernels of the second kind in a 
way  modeled after the split of Legendre functions $P_l$ into Legendre
 functions of the second kind,  $Q_l$,
\ba
&& \mathfrak{V}(\x_0,\tilde{\chi}, \x_2 , \chi_2, \x_1, \chi_1)
= \  \label{kernel2ndKind}\nn  \\ && \frac {\pi}{\sin(l+c)}
[\mathfrak{Q}(\x,\tilde{\chi} ,\x_2,\chi_2,\x_1, \chi_1)-  \nn \\
&& \qquad \qquad - \int d^D\y \Delta^{\tilde{\chi}}(\x,\y )\mathfrak{Q}(\y,\chi,\x_2,\chi_1,\x_1, \chi_2)] 
\nn
\ea
such that the partial Fourier transform \\
$\mathfrak{Q}(p,\tilde{\chi} ,\x_2,\chi_2,\x_1, \chi_1)$\\
{\bf is an entire holomorphic function of $p$.}
\subsection{Continuation to Minkowski apace}
After the split, one can
deform the path of the $c$-integration in the partial wave expansion involving
$g(\chi)$, $\chi=[l,h+c])$, $h=D/2$\\picking up a sum over residues as shown in the following picture\\[4mm]
\setlength{\unitlength}{0.65mm}
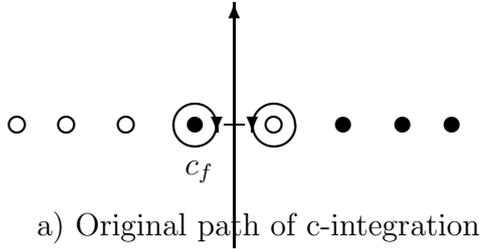
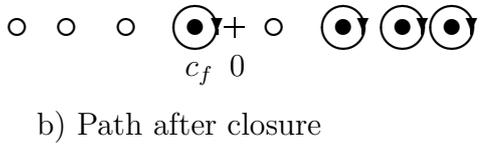
\begin{picture}(100,95)(-2.6,0)
\thinlines
\put(0,50){
\put(38,25){\line(1,0){4}}
\thicklines
\put(40,0){\vector(0,1){50}}
\put(32,25){\circle*{3}}
\put(30,15){$c_f$}
\put(48,25){\circle{3}}
\put(32,25){\circle{9}}
\put(36.4,24.5){\vector(0,-1){1}}
\put(48,25){\circle{9}}
\put(43.6,24.5){\vector(0,-1){1}}

\put(62,25){\circle*{3}}
\put(74,25){\circle*{3}}
\put(84,25){\circle*{3}}

\put(18,25){\circle{3}}
\put(6,25){\circle{3}}
\put(-4,25){\circle{3}}

\put(0,2.5){a) Original path of c-integration} 
}

\put(-100,0){
\thinlines
\put(138,25){\line(1,0){4}}
\put(140,23){\line(0,1){4}}
\thicklines
\put(132,25){\circle*{3}}
\put(130,15){$c_f$}
\put(139,15){$0$}

\put(132,25){\circle{9}}
\put(136.4,24.5){\vector(0,-1){1}}
\put(148,25){\circle{3}}

\put(162,25){\circle*{3}}
\put(162,25){\circle{9}}
\put(166,24.5){\vector(0,-1){1}}

\put(174,25){\circle*{3}}
\put(174,25){\circle{9}}
\put(178,24.5){\vector(0,-1){1}}

\put(184,25){\circle*{3}}
\put(184,25){\circle{9}}
\put(188,24.5){\vector(0,-1){1}}

\put(118,25){\circle{3}}
\put(106,25){\circle{3}}
\put(96,25){\circle{3}}

\put(100,3){b) Path after closure} 
}
\end{picture}


This results in a discrete sum of conformal blocks
as shown before, i.e. the expected OPE.
\begin{figure}[h!]
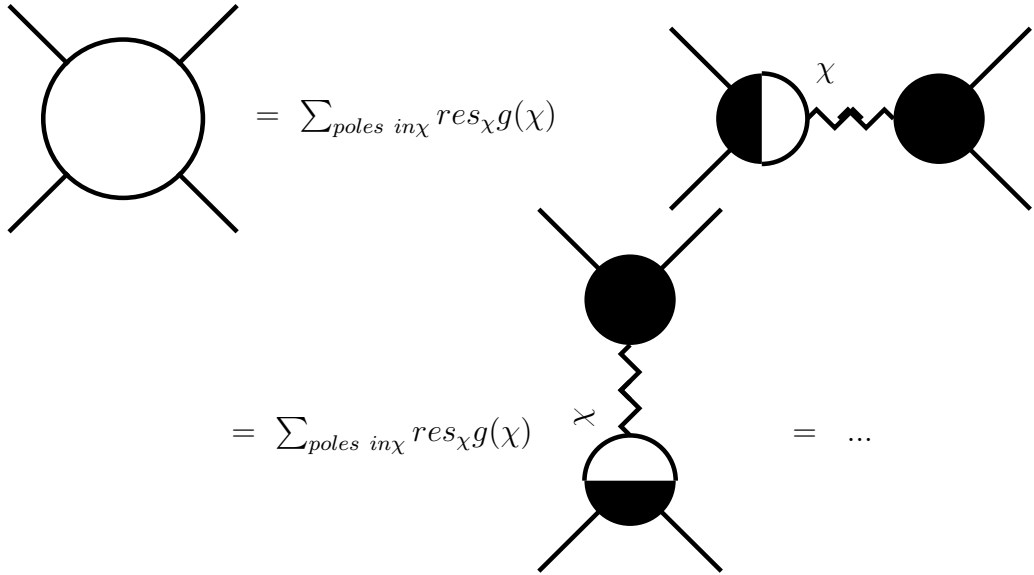

\psset{unit=1.2cm}
\psset{dotsize=0.2 0}
\psset{linewidth=.05}
\def\CircleLeftFoot{
 \psline(-1.,-1.)(-0.3,-0.3)
}
\def\CircleLongLeftFoot{
 \psline(-1.,-1.)(1.,1.)
}
\def\CircleRightFoot{
\psline(1.,-1.)(0.3,-0.3)
}
\def\CircleLongRightFoot{
\psline(1.,-1.)(-1.,1.)
}
\def\CircleFeet{
\psline(-1.,-1.)(-0.3,-0.3)
\psline(1.,-1.)(0.3,-0.3)
}
\def\OvalShort { 
\psline(-.5,-.5)(.5, -.5)
\psline(-.5,.5)(.5,.5)
\psarc(-.5,0){.5}{90}{270}
\psarc(.5,0){.5}{270}{90}
}
\def\OvalLong { 
\psline(-.75,-.5)(.75, -.5)
\psline(-.75,.5)(.75,.5)
\psarc(-.75,0){.5}{90}{270}
\psarc(.75,0){.5}{270}{90}
}
\def\BSKern {
\psline(-.5,-.5)(.5, -.5)
\psline(-.5,.5)(.5,.5)
\psarc(-.5,0){.5}{90}{270}
\psarc(.5,0){.5}{270}{90}
\psline[linestyle=dotted](-1.,0)(1.,0)
}

\def\CircleUp{
\pscircle(0,0){.5}
\psline(0,.5)(0,1.5)
}
\def\OvalShortTwoFeet {
\psline(-1.,-1.)(-0.5,-0.5)
\psline(1.,-1.)(0.5,-0.5)
}

\def\OvalUpFeet {
\psline(-1.,1.)(-.5,.5)
\psline(1,1)(.5,.5)
}

\def\OvalShorFteet {
\psline(-1.,-1.)(-0.5,-0.5)
\psline(1.,-1.)(0.5,-0.5)
}
\def\zigzagOne{
\psline(0,0)(0.1,0.1)(-0.1,0.3)(0,0.4)
}
\def\zigzagTwo{
\psline(0,0)(0.1,0.1)(-0.1,0.3)(0.1,0.5)(0,0.6)
}
\def\CGKern{
\pscircle*(0,0){.5}
\rput(0,0.5){\zigzagTwo}
}

\def\PartialWave{
 \rput{-90}(0,0){
 \CGKern
 \CircleFeet
 }
\rput{90}(2.2,0){
\CGKern
\CircleFeet
 }
}
\def\PartialWaveQ{
 \rput{-90}(0,0){
 \QKern
 \CircleFeet
 }
\rput{90}(2.2,0){
\CGKern
\CircleFeet
}

}
\def\PartialWaveFeetDown{
\rput{-90}(0,0){
\CGKern
  }
 \CircleLongLeftFoot
\rput{90}(2.2,0){
 \CGKern
  }
 \rput(2.0,0){\CircleLongRightFoot}
}
\def\PartialWaveUp{
\rput{90}(0,0){\PartialWave}
}
\def\FourPointFct{
\pscircle(0,0){0.9}
\psline(-1.25,-1.25)(-0.60,-0.60)
\psline(0.60,0.60)(1.25,1.25)
\psline(1.25,-1.25)(0.60,-0.60)
\psline(-0.60,0.60)(-1.25,1.25)
}

\def\FourPointMellin{
\rput(-2.9,0){$=\int d\delta \ M(\{ \delta_{ij}\})$}
\psline(-1.25,-1.25)(-1.25,1.25)
\psline(-1.25,-1.25)(1.25,-1.25)
\psline(-1.25,-1.25)(1.25,1.25)
\psline(1.25,1.25)(-1.25,1.25)
\psline(1.25,1.25)(1.25,-1.25)
\psline(1.25,-1.25)(-1.25,1.25)
}

\def\QKern{
\psarc*(0,0){.5}{90}{-90}
\psarc(0,0){.5}{-90}{90}
\psline(-1.,-1)(-.3,-.3)
\psline(-1,1)(-.3,.3)
\rput(0.5,0){\zigzagTwoHor}
\rput(0.7,0.5){$\chi$}
}

\def\QKerntilde{
\psarc*(0,0){.5}{90}{-90}
\psarc(0,0){.5}{-90}{90}
\psline(-1.,-1)(-.3,-.3)
\psline(-1,1)(-.3,.3)
\rput(0.5,0){\zigzagTwoHor}
\rput(0.7,0.5){$\tilde{\chi}$}
\rput(1.2,0.0){\zigzaglongHor}
}
\def\zigzaglongHor{
\psline(0,0)(0.1,0.1)(0.3,-0.1)(0.5,0.1)(0.7,-0.1)(0.8,0)
}

\def\zigzagTwoHor{
\psline(0,0)(0.1,0.1)(0.3,-0.1)(0.5,0.1)(0.6,0)
}

\def\SplitCGrhs{
 \QKern 
 \rput(1.5,0){$+$}
 \rput(2.7,0){\QKerntilde}
}
\pspicture(0,0)(3,7.0)
\rput(0,5){
 \rput(1,0){\FourPointFct}
 \rput(4.1,0){$=\ \sum_{poles\ in \chi} res_\chi g(\chi)$} 
 \rput(8,0){\QKern}
 \rput(10,0){\rotatebox{90}{\CGKern \CircleFeet} }
}
\rput(6.5,1.0){
 \rotatebox{90}{\QKern}
  \rput(0,2){
   {\rotatebox{180}{\CGKern \CircleFeet}}
 }
}
\rput(5.7,1.5){$=\ \sum_{poles\ in \chi} res_\chi g(\chi) \qquad \qquad \qquad \qquad  = \ \  . . .$}
\endpspicture
\caption{
discrete version of the crossing relation for partial wave $g(\chi)$
}
\end{figure}

\section{\large Recent progress}
\subsection{Conformal Regge theory}
Families of poles of the partial wave $g(\chi=[l, \delta])$ labelled by 
(even or odd) $l$ are analogues of Regge trajectories \cite{Mack6,CostaGoncalvesPenedones,SimmonsDuffin}. But in contrast to conventional Regge theory the pole ``approximation'' is exact!
\subsection{Analysis of the conformal blocks}
Dolan and Osborn \cite{DolanOsborn} determined the conformal blocks for even dimension $D$. Recursion relations were studied by Penedones, Trevisani and Yamazaki \cite{PenedonesTrevisaniYamazaki}

El-Showk, Paulos, Poland, Rychkov, Simmons-Duffin, Vichi 
\cite{ElShowkPaulosPolandRychkovSimmonsDuffinVichi}\\
generalized the result to arbitrary $D$ and use numerical evaluation to 
bound and find critical indices for the {\bf Ising model}

\subsection{Mellin representation}
I proposed to 
 consider bootstrap in Mellin space in 2009 \cite{Mack6}. The Mellin representation was advocated as a natural language by Fitzpatrick, Kaplan, Penedones, Raju and van Rees
\cite{FitzpatrickKaplanPenedones} 
  and further studied by  Dey, Ghosh and Sinha \cite{DeyGhoshSinha}(2018)\\ 
\begin{figure}[t]
\psset{unit=0.6cm}
\psset{dotsize=0.2 0}
\psset{linewidth=.05}
\def\CircleFeet{
\psline(-1.,-1.)(-0.3,-0.3)
\psline(1.,-1.)(0.3,-0.3)
}
\def\OvalShort { 
\psline(-.5,-.5)(.5, -.5)
\psline(-.5,.5)(.5,.5)
\psarc(-.5,0){.5}{90}{270}
\psarc(.5,0){.5}{270}{90}
}
\def\OvalLong { 
\psline(-.75,-.5)(.75, -.5)
\psline(-.75,.5)(.75,.5)
\psarc(-.75,0){.5}{90}{270}
\psarc(.75,0){.5}{270}{90}
}
\def\BSKern {
\psline(-.5,-.5)(.5, -.5)
\psline(-.5,.5)(.5,.5)
\psarc(-.5,0){.5}{90}{270}
\psarc(.5,0){.5}{270}{90}
\psline[linestyle=dotted](-1.,0)(1.,0)
}

\def\CircleUp{
\pscircle(0,0){.5}
\psline(0,.5)(0,1.5)
}
\def\OvalShortTwoFeet {
\psline(-1.,-1.)(-0.5,-0.5)
\psline(1.,-1.)(0.5,-0.5)
}

\def\OvalUpFeet {
\psline(-1.,1.)(-.5,.5)
\psline(1,1)(.5,.5)
}

\def\OvalShorFteet {
\psline(-1.,-1.)(-0.5,-0.5)
\psline(1.,-1.)(0.5,-0.5)
}
\def\zigzagOne{
\psline(0,0)(0.1,0.1)(-0.1,0.3)(0,0.4)
}
\def\zigzagTwo{
\psline(0,0)(0.1,0.1)(-0.1,0.3)(0.1,0.5)(0,0.6)
}
\def\CGKern{
\pscircle*(0,0){.5}
\rput(0,0.5){\zigzagTwo}
}

\def\PartialWave{
\rput{-90}(0,0){
\CGKern
\CircleFeet
 }
\rput{90}(2.2,0){
\CGKern
\CircleFeet
 }
}
\def\FourPointFct{
\pscircle(0,0){0.9}
\psline(-1.25,-1.25)(-0.60,-0.60)
\psline(0.60,0.60)(1.25,1.25)
%
\psline(1.25,-1.25)(0.60,-0.60)
\psline(-0.60,0.60)(-1.25,1.25)
}

\def\FourPointMellin{
\rput(-6,0){$=\int d\delta \ M(\{ \delta_{ij}\})$}
\psline(-1.25,-1.25)(-1.25,1.25)
\psline(-1.25,-1.25)(1.25,-1.25)
\psline(-1.25,-1.25)(1.25,1.25)
\psline(1.25,1.25)(-1.25,1.25)
\psline(1.25,1.25)(1.25,-1.25)
\psline(1.25,-1.25)(-1.25,1.25)
}

\pspicture(-1,0)(15,4.5)


\rput(0,4){
 \rput(0,1.25){\FourPointFct}
 \rput(4,1.25){ $\ = \int d\chi \ g(\chi) $}
 \rput(8.5,1.25){\PartialWave}
 \rput(9.5,1.7){$\chi$}
}
\rput(12,1.25){\FourPointMellin}
\endpspicture
\caption{ Partial wave expansion and Mellin representation for the 4-point function.$<\phi(x_4)...\phi(x_1)>$.  The lines in the Mellin representation represent propagators 
$\Gamma(\delta_{ij}(\xi_i\xi_j)^{-\delta_ij}$ in covariant language
Integration is over $\delta_{ij}=\delta_{ji}$ such that $\sum_{i}\delta_{ij}=d_i$ when the scalar field $\phi(x_i)$ has dimension $d_i$
 \label{fig:partialWaveAndMellin.tex}}
\end{figure} 

%
 The lines in the Mellin representation represent propagators 
$\Gamma(\delta_{ij})(\xi_i\xi_j)^{-\delta_{ij}}$ 
 in Dirac's covariant language.
Integration is over $\delta_{ij}=\delta_{ji}$  such that
 $\sum_{i}\delta_{ij}=d_i$  when the scalar field $\phi(x_i)$
 has dimension $d_i$
%
\subsection{Pismaks simplified version of the crossing relation}
Pismak \cite{Pismak} proposed to study a simplified version of the crossing relation for the partial wave $g(\chi)$ as shown in Figure 5.
\begin{figure}[h]
\psset{unit=1cm}
\psset{dotsize=0.2 0}
\psset{linewidth=.05}
\def\CircleRightFoot{
\psline(1.,-1.)(0.3,-0.3)
}
\def\CircleFeet{
\psline(-1.,-1.)(-0.3,-0.3)
\psline(1.,-1.)(0.3,-0.3)
}
\def\CircleWLeftFootLong
{
\pscircle(0,0){.5}
\psline(1.7,-1.7)(0.3,-0.3)
}
\def\CircleWRightFootLong
{
\pscircle(0,0){.5}
\psline(-1.7,-1.7)(-0.3,-0.3)
}

\def\CircleWFeetLong
{
\pscircle(0,0){.5}
\psline(-1.7,-1.7)(-0.3,-0.3)
\psline(1.7,-1.7)(0.3,-0.3)
}
\def\zigzagOne{
\psline(0,0)(0.1,0.1)(-0.1,0.3)(0,0.4)
}
\def\zigzagTwo{
\psline(0,0)(0.1,0.1)(-0.1,0.3)(0.1,0.5)(0,0.6)
}
\def\CGKern{
\pscircle*(0,0){.5}
\rput(0,0.5){\zigzagTwo}
}
\def\PartialWave{
 \rput{-90}(0,0){
 \CGKern
 \CircleFeet
  }
 \rput{90}(2.2,0){
 \CGKern
\CircleFeet
 }
}

\def\PartialWaveHorizontalBottomLong{
\rput(1.15,.5){$\chi$}
 \rput{-90}(0,0){
 \CGKern
\CircleWLeftFootLong 
 }
 \rput{90}(2.2,0){
 \CGKern
\CircleWRightFootLong
 }
\rput(-1.7,-2.1){$3$}
\rput(3.9,-2.1){$4$}
}
\def\PartialWaveUp{
\rput{90}(0,0){
 \rput{-90}(0,0){
 \CGKern
 \CircleWFeetLong
  }
 \rput{90}(2.2,0){
 \CGKern
\CircleRightFoot
  }
 }
\rput(-0.9,3.55){$p_1=0$}
%
\psline[linearc=.25](1.7,-1.65)(1.0,3.0)(.3,2.55)
\rput(1.5,2.0){$m$}
\rput(1.5,1.5){$\alpha$}
\rput(0.65,1.0){$\chi^\prime$}
\psline(1.4,1.75)(1.1,1.75)
\rput(-2,0.5){$= \int g(\chi^\prime)$}
}

%
\def\PartialWaveHorizontalBottomLongP10{
\PartialWaveHorizontalBottomLong
\psline(-.3,.3)(-1.,1)
\rput(-1.3,1.3){$p_1=0$}
\rput(2.2,0){
\psline[linearc=.25](.3,.3)(1.,1.)(1.7,-1.7)
\psline(1.1,0.0)(1.4,0.0)
\rput(1.6,.3){$m$}
\rput(1.6,-.4){$\alpha$}
 }
}
\pspicture(-1,0)(15,5.0)
\rput(10,1.5){
\PartialWaveUp
}
\rput(2,2,0.5){
\rput(-1.5,0){$\int  g(\chi)$} 
\PartialWaveHorizontalBottomLongP10
}
\endpspicture
\caption{ Pismaks version of the crossing symmetry of the partial wave $g(\chi)$. In the Ising model with fields $\phi(x)$ and $\phi^2(x)$, $g(\chi)$
is a $2\times 2$ matrix.
 \label{fig:crossingKernelPismak}
}
\end{figure}

%
\subsection{Spacetime derivation of the Lorentzian OPE inversion formula}
Caron-Huot \cite{CaronHuot}, and Simmons-Duffin, Stanford, Witten \cite{SimmonsDuffinStanfordWitten}
 derive a formula for the partial wave 
$g(\chi=[\Delta, J] ) = n_{\Delta,J}I_{\Delta,J}$  expressed in terms of vacuum 
vacuum expectation value of commutators of the  four scalar fields
 $\phi_i(x_i)$ of dimension $d_i$ in Minkowski space,  $n_{\Delta,J}$
is  an explicitly given normalization factor
\ba
I_{\Delta,J} &=& -\hat{C}_J(1)
\left[ \int_{3>1,2>4} \frac {d^D x_{3}d^Dx_{4}}{vol(SO(D-1)}\right.\nn \\
&&
 \frac{ <\Omega, [\phi_4(x_4),\phi_2(x_2)][\phi_1(x_1),\phi_3(x_3)]]|\Omega>}{ |x_{34}|^{J+2D-d_3-d_4-\Delta}}
\nn \\
&&
\left.
(m\cdot x_{34})^J \theta(m\cdot x_{34}) \right.\nn \\
 &+& (-1)^J  \int _{4>1,2>3}
  \frac{d^D x_3 d^D x_4}{vol(SO(D-1))}\nn \\
&&
 \frac{<\Omega|[\phi_3,(x_3),\phi_2(x_2)][\phi_1(x_1),\phi_4(x_4)]|\Omega>}{|x_{34}|^{J+2D-d_3-d_4-\Delta}}
\nn \\
&& \left.
 (-m\cdot x_{34})^J  \theta(-m\cdot x_{34}) 
\right]
\nn 
\ea
where $m$ is the nullvector $m^\mu=(1,1,0,...,0)$ where the second component is the time direction, and coordinates $x_1,x_2$ 
have been fixed to 
$x_1=(1,0,0,...), $, $x_2=0$. The notation $i>j$ means that $x_i$ is in the future light cone of $x_j$.

An explicit expression for 
 $\hat{C}_J(1)$  is given.

The formula has the following advantages: It can be {\bf analytically continued in the spin $J$}, and for real dimension and spin, the integrand satisfies {\bf positivity conditions}.
%

\bibliographystyle{plain}

\end{document}